\documentstyle[12pt,epsf]{article}

\def\case#1/#2{\textstyle\frac{#1}{#2} }
\begin{document}

\title{Caustics of Compensated Spherical Lens Models}
\author{G. F. R. Ellis $^{1}$, {\sc D. M. Solomons $^{1}$} \\
{\normalsize {1. {\it Department of Applied Mathematics, University of Cape
Town,}}}\\
{\normalsize {\it Rondebosch 7700, Cape Town, South Africa.}}}
\maketitle

\begin{abstract}
We consider compensated spherical lens models and the caustic surfaces
they create in the past light cone. 
Examination of cusp and crossover angles associated
with particular source and lens redshifts gives explicit lensing
models that confirm the claims of 
Paper I \cite{bi:I}, namely that area distances can differ by
substantial factors from angular diameter distances even when averaged
over large angular scales. `Shrinking' in apparent sizes occurs, 
typically by a factor of 3 for a single spherical lens, on the scale
of the cusp caused by the lens; summing over many lenses
will still leave a residual effect.
\end{abstract}

\vspace*{0.2truecm}

\begin{center}
{\it Subject headings:\\cosmology\,-\,gravitational lensing,
\,-\,observational tests}
\end{center}

\newpage
\noindent
\section{Introduction}
\label{sec: Intro} 

This paper continues the study started in Paper I \cite{bi:I}
of how area distances behave in universes where strong gravitational 
lensing takes place. That paper considered the claim \cite{bi:weinberg} 
that although individual lensing
masses alter area distances for ray bundles that pass near by them, photon
conservation guarantees the same area distance-redshift relation as in exact 
It was shown in \cite{bi:I} that this claimed compensation result is 
incorrect once
one has passed caustics, which are necessarily the result of strong
gravitational lensing; consequently (by continuity) the result is not true
in general. Indeed it has to be wrong because area distances are determined
by the gravitational field equations (essentially through the null
Raychaudhuri equation) quite independently of the issue of photon
conservation (which is determined by Maxwell's equations and is valid
whatever the space-time curvature). Thus the latter cannot causally
determine area distances. In fact at large distances, `shrinking' takes
place in that distant areas subtend smaller solid angles than they would in
a FL universe model; and this effect will remain even when the observations
relate to large angles. The way these small effects for individual lenses add
up to give a significant averaged effect over the whole sky is discussed in
Paper I. This may affect high-redshift number counts and Cosmic
Background Radiation (`CBR') anisotropy observations at very small angles.

The general argument has been given in I. Specific spherically symmetric
examples (viewed from the centre, and so without caustics) 
are presented in \cite{bi:II}, which thereby gives a rigorous proof 
of the existence of the effect we claim; but the models used are unrealistic
as models of the real universe. In this Paper (Paper II), we consider 
exact examples with caustics displaying the shrinking effects discussed 
in Paper I. We show how to
calculate the magnitude of the effect analytically and numerically for
single spherically symmetric compensated lenses to which we can apply 
the thin-lens approximation, and look in detail at `top hat' lenses, which 
are the simplest in this class. We refer to and mainly follow the notation 
of Schneider, Ehlers and Falco (1992) \cite{bi:Schneider92} 
(`SEF').


\section{The Gravitational Lensing Equations}

\label{sec:lensing} 
We consider compensated spherically symmetric lenses in an Einstein-de Sitter
background universe. The effect of the lens will be represented
by the usual thin lens
approximation, and we use the scaled variables of SEF. 

\subsubsection{Background Relations}
The angular diameter distances between the observer and lens,
observer and source, and lens and source in this background
are $D_d$, $D_s$ and $D_{ds}$ respectively. In an Einstein - De Sitter 
model  ($\Omega = 1$, no clumping),
\begin{equation}
D_{ds} = \frac{2c}{H_0}\frac{(1+z_d)^{1/2}\, (1+z_s) - (1+z_d)\,
(1+z_s)^{1/2}}{(1+z_d)\, (1+z_s)^2}\; \label{eqn:20}
\end{equation}
[see SEF (4.57)], and $D_d$ (respectively, $D_s$) is obtainable from $D_{ds}$
by setting $z_d \rightarrow 0, z_s \rightarrow z_d$ 
(respectively, $z_d \rightarrow 0$).  
follows, the dimensionless ratio 
${\cal R} = {\frac{H_0 }{2c}} {\frac{D_d D_{ds} }{D_s}}  $
is important. For a given lens position $z_d$, as $z_s \rightarrow \infty$
 this has the limiting value 
$
{\cal R}_\infty = \frac{(1+z_d)^{1/2} - 1}{(1+z_d)^2}\; ,
$
which has a maximum value of ${\frac{2 }{3}} {\frac{9^2 }{16^2}} = 0.21 $ 
when $z_d = 9/7$.

If a source in a FLRW universe with scale factor $a(t)$
emits a signal at time $t_s$ which is received at time $t_0$, 
then the proper distance at time $t_s$ between the
source and observer is
$\ell = a(t_s) \int^{t_0}_{t_s}\frac{dt}{a(t)} \,.$
For an Einstein-de Sitter universe $a(t) = 
t^{2/3}$, the Hubble parameter is $H(t) = (2/3) t^{-1}$, 
and $1+z = t_0^{2/3}/t_s^{2/3}$, so 
\begin{equation}
\ell = 3c t_s^{2/3} (t_0^{1/3} - t_s^{1/3}) = (2c /H_0) 
(1+z)^{-3/2} (\sqrt{1+z
}\,-\,1)  \label{eqn:16}
\end{equation}
is this distance in the background universe.

\subsection{Compensated lenses}

If the energy density is $\rho(\vec{x})$ the fractional matter perturbation 
$\delta(\vec{x})$ in an inhomogeneity is related to the matter source by 
\begin{equation}
\delta(\vec{x}) = \frac{\rho(\vec{ x}) - \rho_0}{\rho_0} ~~\Leftrightarrow~~
\rho(\vec{x}) = \rho_0 \left(\delta(\vec{x}) + 1\right)  
\; .  \label{eqn:4}
\end{equation}
where $\rho_0$ is the average energy density over a hypersurface of 
constant time ${\cal S}$, defined by
\begin{equation}
\int_{{\cal S}} \rho(\vec{x})d^3x = \rho_0 \; .  \label{eqn:3}
\end{equation}
Integrating (\ref{eqn:4}) over ${\cal S}$, 
$
\int_{{\cal S}}\delta(\vec{x})d^3x = \frac{1}{\rho_0}\int_{{\cal S}}\left[
\rho(\vec{x}) - \rho_0\right]d^3x\; ; $
so by (\ref{eqn:3}), 
\begin{equation}
\int_{{\cal S}} \delta(\vec{x})d^3x = 0\; ,  \label{eqn:5}
\end{equation}
which is the condition for a compensated perturbation that has been formed
by rearrangement of matter in a uniformly distributed background
with matter density $\rho_0$.
Equivalently, the density $\rho$ averages out to the correct 
background value $\rho_0$;
if this is not true, then the background density has been wrongly assigned 
\cite{bi:Elljak99}. Clearly this means that $\delta(\vec{x})$ 
positive. By (\ref{eqn:4}), no negative densities will occur iff 
\begin{equation}
\rho_0 > 0\,, ~~ \delta(x) > -1 ~{\rm everywhere}.  \label{eq:pos}
\end{equation}

We will assume these conditions to be true for the matter inhomogeneities
causing lensing. Then in the lensing equations that follow, the
quantities and relations will all refer to the variation from what they
would have been in the background model (i.e. if there had been no lens). 
Thus the lensing
`surface mass density' $\sigma$ will mean the projected surface mass density
in the lens plane
arising from the density difference $\delta\rho = \rho - \rho_0$ 
from the background 
value, which will be chosen so that the compensation condition
(\ref{eqn:5}) is true; and the `bending angle' 
will be the deviation in direction at the lens from
what it would have been in the background model. This will be given 
via the usual thin-lens equations, with the surface mass density as 
just defined.

\subsubsection{Simple compensated lenses}

We define a {\it simple compensated lens} (`SCL') to be a spherically
symmetric compensated lens where $\delta(\vec{x}) = \delta(|\vec{x}|)$ is
positive for an inner domain $0 \leq |\vec{x}| < 1$, negative for an outer
domain $1 < |\vec{x}| < \lambda $, $(\lambda > 1)$, and zero at larger
radii, i.e. for $\lambda < |\vec{x}|$. This configuration would naturally
arise by formation of a spherical massive object through gathering together
material from an initially uniform substratum. In the sequel we consider a
particularly simple form of SCL, namely a top hat lens.

\subsection{The lensing equation in scaled variables}
Given a choice of length scale $\xi_0$ in the lens plane, there is a
corresponding length scale $\eta_0 = \frac{D_s }{D_d}\xi_0$ in the source
plane. From the position vector $\vec{\eta}$ of the source relative to the 
optic axis in the source plane, and the impact vector 
$\vec{\xi}= D_d \vec{\theta} $ in the lens plane, 
where (vector) $\vec{\theta}$ is the observational angle from the optical axis, 
we define corresponding scaled variables $\vec{x}$, $\vec{y}$ by
\begin{equation}
\vec{x} = \frac{\vec{\xi}} {\xi_0},~~\vec{y} = \frac{\vec{\eta}}
{\eta_0}\end{equation}
 [SEF (3.5)]. The surface mass density $\sigma(|\vec{\xi}|)$
for thin spherical lenses can be rescaled as 
\begin{equation}
\kappa(x) = \frac{\sigma(\xi_0 x) }{\sigma_{cr}}\,,~~ \sigma_{cr} : = {
\frac{c^2 }{4 \pi G}} {\frac{H_0 }{2c{\cal R}}}\,,~~x = |\vec{x}|\,.  
\label{eqn:sc}
\end{equation}
(SEF 5.4, 5.5). The lens equation can then be written in the
very simple dimensionless form 
\begin{equation}
\vec{y} ~=~ \vec{x} - \vec{\alpha}(\vec{x})  
\label{eqn:lens}
\end{equation}
(SEF 5.6, 8.6) where the scaled (vector) deflection angle $\vec{\alpha}$ is 
related to the true (vector) deflection angle $\vec{\hat{\alpha}}$ by
\begin{equation}
\vec{\alpha}(\vec{x}) = {\frac{2c{\cal R} }{H_0\xi_0}} \vec{\hat{\alpha}} 
(\xi_0 \vec{x})  \label{eqn:10b}
\end{equation}
(SEF 5.7). Because of the spherical symmetry, the deflection is radially
inward and of magnitude $y = |\vec{y}|$ given by 
\begin{equation}
y ~=~ x - \alpha(x) ~=~ x - {\frac{m(x) }{x}}  \label{eqn:sc1}
\end{equation}
where 
\begin{equation}
m(x) = 2 \int_0^x x^{\prime}dx^{\prime}\kappa(x^{\prime})  \label{eqn:m}
\end{equation}
is the dimensionless mass $m(x)$ within a circle of radius $x$ (SEF 8.3);
its first derivative is the dimensionless surface density (SEF 8.13): 
\begin{equation}
m^{\prime}= \frac{dm(r)}{dr} = 2r\kappa(r)\, .  \label{eqn:34}
\end{equation}
Also the change $\Delta \ell$ in radial distance traveled by light
in a given time, as measured at the source, is equal to the
time delay caused by lensing [SEF (4.67), (5.45)]\footnote{The time 
change calculated
in these equations is at most a first order quantity, and so the change 
in radial 
distance travelled can be found from it to first order by calculating 
distance 
as if light travels on null geodesics in the background geometry.}, 
and can be rescaled to 
\begin{equation}
Z = \frac{1 + z_s}{1 + z_d} \left( \frac{2c{\cal R}}{H_0\xi_0^2} \Delta
\ell \right) \,
\end{equation}
[SEF (5.11) and following]. This is given by 
\begin{equation}
Z = \frac{1 }{2} |\, \vec{\alpha}\, |^2 - \psi(x)\; ,  
\label{eqn:24b}
\end{equation}
where $\vec{\alpha}$ is given by (\ref{eqn:10b}) 
and the (rescaled) deflection potential is 
\begin{equation}
\psi(x) = 2 \int_0^x {x}^{\prime} \kappa (x^{\prime}) 
ln \left( \frac{x}{{x}^{\prime}}\right) dx^\prime ~~\Rightarrow~~ 
\alpha(x) = \frac{d\psi(x)}{dx}\,.
\end{equation}
[SEF (8.7)-(8.9)]. 

\subsubsection{The shrinking ratios}
Finally, pointwise over the sky, the angular
shrinking factor $\gamma$ which relates observed distances corresponding
to a given angle in the real lumpy
universe to those in the background smoothed-out universe (see Paper I) is
\begin{equation}
\gamma = |dy/dx|
\end{equation}
which can be averaged over a stated angle $\Delta\Theta$ to give
the average angular shrinking factor $\langle \gamma \rangle$ over that
angle. Similarly the pointwise area shrinking factor $\beta$ 
which relates observed areas corresponding to a given solid angle
in the real lumpy
universe to those in the background smoothed-out universe (see Paper I) is 
\begin{equation}
\beta = |\det\,J\,|,~~ J =\,\left|\, {\frac{\partial\, \vec{y} }{\partial\, 
\vec{x}}}\, \right|.
\label{eqn:det}
\end{equation}
This can be averaged over a solid angle $\Delta \Omega$ to give the
average area shrinking factor $\langle \beta \rangle$ over that solid angle.

\subsubsection{The overall effect}

Together the radial and transverse equations (\ref{eqn:24b}),
(\ref{eqn:lens})
 give the deflection of each
null ray relative to the background geometry, and hence the shape of the
perturbed light cone in the real lumpy space-time. These deflections are not
independent: they are related by the fact that the speed of light is locally
unity, so that the actual light path is stationary w.r.t. variation of the
arrival time delay. Consequently a sideways deflection (which increases the
distance to be traveled) is compensated by an inwards deflection (reducing
the distance to be traveled), so the (tangential) lensing equation is a
consequence of the (radial) time delay equation (SEF pp.146-147, 170-171).
It is this combination of radial and tangential effects, implied by the
above equations, that gives the light cone caustics their characteristic
shapes (see Figure 1). 


\section{Angles and Distances}

\label{sss: CAD}

Consider now angles and distances in the perturbed space-time. We start with
angular diameter distance. Consider the source plane of a lens in
direction $(\vec{\theta})$. Image points of nearby directions
will be displaced from their background position $(\ell,\vec{x})$ 
by the (scaled) displacement $(\Delta\ell,\Delta \vec{\eta})$ 
$(Z,\vec{y}- \vec{x}) = (Z,- \vec{\alpha})$
where the first part is the radial component of the displacement, 
given by (\ref{eqn:24b}), 
and the second is the tangential component (in the source plane)
 given by (\ref{eqn:sc1}). 
If we move our
viewing direction through an arc in the sky, the image point will move; for
simplicity we will consider an arc where only one angular component $\theta$
only varies. This gives a 2-dimensional section of the full 3-dimensional 

As we vary $\theta$ through $d\theta$, the (scaled) tangential distance 
traveled will be $dy^2$, and the radial change of distance $dZ^2$
will be much less than this.
Thus the total {\it distance traveled} $DT$ due to an
angle increase $\Delta\theta$ is, to good approximation, 
\begin{equation}
DT(\Delta\theta) = \int_{\Delta\theta} \, \left|{\frac{\partial y (\theta)}
{\partial \theta}}\right| d\theta\; ,  
\label{eqn:25k}
\end{equation}
where we sum all distances with a positive sign, thus determining the
total increment in $|y|$ (see Paper 1).
In terms of normalised magnitudes when spherical lensing takes place, the
integrand is just $\left|\partial y/\partial \theta\right| = 
|1 - d\alpha/dx| \partial x/\partial \theta.$ 
By contrast, the background distance is the same expression but with
integrand $|\partial x/\partial \theta|$, and  
{\it Distance gained} $DG$ is the distance moved from the starting point: 
\begin{equation}
DG(\Delta\theta) = \int_{\Delta\theta} \, 
{\frac{\partial y (\theta) }{\partial \theta
}} \,d\theta ~=~ y(\theta) - y(0)\,.  \label{eqn:25f}
\end{equation}
In this case we subtract off those regions where $\partial y/\partial \theta$
is negative, ending up simply with the increment of $y$.

Now when ${\frac{\partial y (\theta) }{\partial \theta}}$ is positive,
distance traveled is the same as distance gained. However when ${\frac{\partial
y (\theta) }{\partial \theta}}$ changes sign, we have cusps forming
(see Paper 1) and in the formula (\ref{eqn:25k}) for distance traveled 
the integral is over the curve corresponding to all values of $
\theta$ and hence traverses the cusp backwards and forwards, see Figure 1. 
This is different from distance gained; the latter is then given by 
the integral (\ref{eqn:25k}) but where now 
the integral is over the (connected) curve $\gamma$ excluding the cusp 
sections, so
that ${\frac{\partial y (\theta) }{\partial \theta}}$ is positive over
all the curve traversed.

The {\em Change in Distance Gained} due to the presence of the deflector 
is small in all cases. The effect for
large angular scales does not average to zero when we have a distribution of
lenses, but it is very small (the change from the background value is given
by the difference of $- D_{ds}\; \vec{\alpha}\,(\vec{\xi})$ at the two ends,
corresponding to a minute of arc at most). 
The {\it Change in Distance Traveled} 
$\Delta DT$ is given by the integral (\ref{eqn:25k}), 
but now taken over all the closed
loops $\gamma_c$ that are excluded when one calculates distance gained. 
The effect at each lens is small, but it is cumulative. Hence when there are
a large number of lenses, the effect can be large (as discussed in Paper I).

We are also interested in the true {\em Cosmological Area Distance} and so 
the question, as we look over a given solid angle, what area does that cover
at the source? This change is given pointwise by the determinant of the lens
equation, see (\ref{eqn:det}). We give explicit
expressions for this determinant in the following paragraphs. A radial
increase of size will be partly compensated by a transverse decrease of
size, see e.f. Gunn and Press \cite{bi:gp}), so the area distance will not
relate very simply to the (radial) angular size distance.

We see then that before caustics form, distance traveled and distance
gained are both very similar (and very close to the background value, on
large angular scales). Thereafter, they can be very different (as was argued
in Paper I). To calculate this, we must locate the cusps and caustics.

\subsection{Caustics and Critical Curves}

\label{sec:Curves} {\it Caustics} in a source plane are the points in the
plane where the Jacobean of the lensing map is singular. {\it Critical curves
} are the points in the lensing plane where the light rays pass that will
end up at caustics at the source plane. They can be located by determining
the zeros of the Jacobean of the lensing map. The set of caustic points in
space-time for all source planes form the space-time caustic set.

\subsubsection{The Jacobean}

Considering a spherical lens ${\bf \hat{L}}$ 
centered at the origin of the
Cartesian $XY$-plane, the Jacobian matrix $J\,= \left(\, \frac{\partial\, 
\vec{y}}{\partial\, \vec{x}}\, \right)$ of the (transverse) lens mapping in
the lens plane has determinant 
\begin{equation}
det\, J = \left(1 - \frac{\alpha(x)}{x} \right)\, \left(1 - \frac{d\alpha(x)}
{dx}\right)\; .  \label{eqn:38}
\end{equation}
(SEF 8.16) which vanishes where either the first or the second brackets on
the r.h.s. vanishes.

When the first bracket in (\ref{eqn:38}) vanishes, the radius $x$ is $x_c$
such that 
\begin{equation}
\frac{m(x_c)}{x_c^2} = 1 ~~\Leftrightarrow ~~ \frac{\alpha(x_c)}{x_c} = 1
~~\Leftrightarrow ~~ \frac{\hat{\alpha}(r_c)}{r_c} = \frac{1}{{\cal R}}\; .
\label{eqn:39}
\end{equation}
Such a critical point occurs for example at ${\bf r} = (x_c,0)$; then there
is a tangent vector $\Xi_t = (0,1)$ to the critical curve at this point, and
since the curve is tangential, $\Xi_t$ is an eigenvector with zero
eigenvalue. Since the tangential critical curves are mapped onto the point 
$\eta = 0$ in the source plane, there exists a caustic there which
degenerates to a single point. The equation of the tangential critical curve
in {\em two dimensions} in the lens plane is then simply $x^2 = x^2_c$,
where $x_c$ solves (\ref{eqn:39}). This corresponds to an Einstein ring
[many images, in a circle, of one point in the source plane].

The determinant $det \, J$ in (\ref{eqn:38}) also vanishes where the last 
\begin{equation}
\frac{d\alpha(x_d)}{dx} = 1\; .  \label{eqn:42}
\end{equation}
This equation describes radial critical curves. Again it corresponds to a
circle in the lens plane. It has a radial eigenvector $\Xi_r$ with
eigenvalue zero. For instance, at ${\bf r} = (\xi,0)$, $\Xi_r = (1,0)$. We
see in the next section it corresponds to a caustic in the source plane [and
a cusp in the surface of constant distance].

\subsubsection{The Cross-over and Cusp Angles}
\label{sec: Fold} The lensing equation (\ref{eqn:lens}) is a two-dimensional
vector equation with (transverse) components $y_1$ and $y_2$ while
the radial equation gives the 3rd-component for the 2-d section of the past
null cone in any surface of constant time. Lensing is radially 
inward with radial displacement magnitude given by (\ref{eqn:sc1}).
The first term on the right is the position that would
have been with no lens; the second term is the effect of the lens.

To express this in terms of the observational angle $\theta$
from the optical axis, we note from
the relations $\vec{\xi}= D_d \vec{\theta} $,
$\vec{x} = \vec{\xi} / \xi_0$ that 
$\vec{x} = D_d \vec{\theta} / \xi_0$. Hence 
the magnitude equation takes the form
\begin{equation}
y (\theta) =   \frac{D_d }{\xi_0} \left(\theta  
    - \frac{D_{ds}}{D_s} \; \hat{\alpha}\,(D_d \theta)\right) .  
\label{eqn:F2}
\end{equation}
(SEF 4.47b, 5.34). The cusp angles $\theta_1$ and $\theta_{-1}$ are 
determined by 
\begin{equation}
{\frac{\partial y}{\partial \theta}}|_{\theta_1} = 0, ~~ {\frac{\partial
y}{\partial \theta}}|_{\theta_{-1}} = 0,  \label{eqn:F6}
\end{equation}
where again by the symmetry, $\theta_1 = - \theta_{-1}$. Differentiating (\ref
{eqn:F2}), this occurs when 
\begin{equation}
0 = D_s - D_{ds} {\frac{\partial \hat{\alpha} }{\partial \xi}} (D_d
\theta_1) D_d  \label{eqn:F7}
\end{equation}
that is 
\begin{equation}
{\frac{\partial \hat{\alpha} }{\partial \xi}} (D_d \theta_1) = {\frac{1 }{
{\cal R}}}  \label{eqn:F8}
\end{equation}
determines the cusp angle $\theta_1$. These angles correspond to the radial
critical points in equation (\ref{eqn:42}). In terms of the bending angle
diagram (Figure 2), this occurs where the curves $y - x = y_1$ are tangent
to the curve $\alpha(x)$. The cusp physical size is 
$ D_s \theta_1; $
 \,twice this distance is the difference between distance gained and distance
traveled, to good approximation. 
 
$\theta_2$ and $\theta_{-2}$ are related by 
\begin{equation}
y(\theta_2) = y(\theta_{-2}) = 0,  \label{eqn:F3}
\end{equation}
where by the spherical symmetry $\theta_2 = - \theta_{-2}$
and the self-intersection of the light cone (given by the first equality
in this equation) occurs on the central line through the lens (as implied by
the second equality). Thus we have from (
\ref{eqn:F2}) 

\begin{equation}
\theta_{-2} = {\frac{D_{ds} }{D_s}} \hat{\alpha}(D_d \theta_{-2})
\label{eqn:F5}
\end{equation}
determines the cross-over angle $\theta_{-2}$. Thus the cross-over angles $
\theta_2$ and $\theta_{-2}$ correspond to the critical points satisfying
equation (\ref{eqn:39}). In terms of the bending angle diagram (see Figure
2) this occurs where the line $y - x = 0$ intersects the curve $\alpha(x)$.

An angle $\theta_{3}$ and corresponding impact parameter $x_3$
yields the same image position as the cusp angle, on the
other side of the caustic: $y(x_3)=-y(x_1)$, and it is {\em this }angle that
we treat as the cut-off in the caustic size. Henceforth, we refer to this as
the {\em cut-off angle }$\theta_{3}$ (and the cut-off on the other side
occurs at $\theta_{-3} = - \theta_3$). 

Finally, the maximum deflection caused by
the lens occurs when $\theta = \pm \theta_m$, where 
\begin{equation}
{\frac{\partial \hat{\alpha} }{\partial \xi}} (D_d \theta_m) = 0
\label{eqn:F10}
\end{equation}
This does not correspond to either of the other angles; indeed it lies
between them. For a SCL centred at $\theta = 0$, if cusps and cross-overs
occur then generically 
\[
0 < \theta_1 < \theta_m < \theta_2 < \theta_3\,. 
\]

The two-dimensional picture obtained by suppressing one angular coordinate
is as shown in Figure 1 (with one radial coordinate and one angular
coordinate). Going to the full 3-dimensional picture, at the source plane,
the whole picture is circularly symmetric about the optical axis at $\theta
= 0$. The cross-over angles at $\theta = \pm \theta_2$ correspond to a
circle in the lens plane but a point (a degenerate caustic) in the source
plane; the cusp angles $\theta = \pm \theta_1$ correspond to circles in both
planes. 

We now apply the preceding theory to Top Hat models.
\section{The Top-Hat Matter Distribution}
\label{sec:TH}  
density $\delta_+$ for $0 \leq |\vec{ x}| < 1$ and a constant outer
density  $\delta_-$ for $1 < |\vec{ x}| < \lambda$ with $\lambda > 1.$
Then the
compensation condition (\ref{eqn:5}) is 
\begin{equation}
\delta_- = - (\lambda^3-1)^{-1}\delta_+ \,,
\end{equation}
Unless otherwise stated, we will assume that $\delta_+$ is positive (so $
\delta_-$ is negative). Then the positivity condition (\ref{eq:pos}) demands
that 
\begin{equation}
0 < \delta_+ < (\lambda^3-1) ~~\Leftrightarrow~~0 > \delta_- > -1\,,
\end{equation}
using the scaled variables, and $\kappa(x)$ will take the form 
\begin{eqnarray}
\kappa(x) = & C \left(\lambda^3\sqrt{1 - x^2} - \sqrt{\lambda^2 - x^2}\,
\right)\,, ~~& ~~0 \leq x \leq 1\,, \\
\kappa(x) = & - C \sqrt{\lambda^2 - x^2}\,, ~~& ~~ 1\leq x \leq \lambda \,,
\\
\kappa(x) = & 0\,, ~~& ~~x > \lambda
\end{eqnarray}
where $C = - 2\rho_0{} \delta_-/\sigma_{cr}$, with a central value $
\kappa(0) = C \lambda (\lambda^2-1) > 0$ and a junction value of $\kappa(1)
= - C \sqrt{\lambda^2-1} < 0$. The surface density will positive for $0 \leq
x < x_+ < 1$, negative for $x_+ < x < \lambda $, and zero for $\lambda < x$,
where 
\begin{equation}
x_+ = \lambda \sqrt{(\lambda^4-1)/(\lambda^6 -1)}\,< 1\,.
\end{equation}

Substituting into (\ref{eqn:sc}) and integrating (\ref{eqn:m}) to find the
mass function $m(x)$, we obtain the following: 
\begin{equation}
m(x) = A f(x), ~~A = - {\frac{4\rho_0 }{3}}\delta_- {\frac{\xi_0 \lambda^3 }{
\sigma_{cr}}} \,,  \label{eqn:m1}
\end{equation}
where the function $f(x)$ is given by 
\begin{eqnarray}
f(x) =& \left(1 - {\frac{x^2 }{\lambda^2}}\right)^{3/2} - \left(1 - x^2
\right)^{3/2}, ~&~0 \leq x \leq 1 \,,  \label{eqn:m2} \\
f(x) =& \left(1 - {\frac{x^2 }{\lambda^2}}\right)^{3/2},~&~ 1 \leq x \leq
\lambda \,,  \label{eqn:m3} \\
f(x) =& 0, ~&~ \lambda \leq x \,.  \label{eqn:m4}
\end{eqnarray}
The function $f(x)$ is a continuous positive even function, with $f(0) =
df/dx(0) = 0$, a single maximum value of $f(x_m) =
(\lambda^2-1)^{3/2}/(\lambda^6-1)^{1/2}$ at $x_m< 1$ given by $x_m^2 =
\lambda^2 (\lambda^4-1)/(\lambda^6 -1) = x_+^2$, and junction values $f(1) =
(\lambda^2-1)^{3/2}/\lambda^3$, $f(\lambda ) = 0 = df/dx(\lambda )$. Near
zero it has the form 
\begin{equation}
f(x) = {\frac{3 }{2}} {\frac{(\lambda^2-1)}{\lambda^2}} x^2 - {\frac{3 }{8}}
\left({\frac{\lambda^4-1 }{\lambda^4}}\right)x^4 + O(x^6)\,.  \label{eqn:exp}
\end{equation}
It follows that $m(x)$ is a continuous non-negative function with $m(0) = 0$ 
and junction values $m(\lambda) = 0$ and $m(1) = A
(\lambda^2 - 1)^{3/2}/\lambda^3$. Its maximum value is at $x = x_m = x_+$,
where $m(x_m) = A f(x_m)$.

Consequently, because any SCL lens can be built up by a superposition of a
sufficient number of top hat lenses, we see that the effective surface
deflection mass $M(r)$ is always positive and is exactly zero at the outer
edge of the compensating region, that is {\it the effective 2-dimensional
surface density $\sigma$ is exactly compensated if the 3-dimensional
fractional density $\delta$ is precisely compensated}. Hence there is no
long-range effect due to the lens: precisely because it is correctly
compensated, the deflection angle $\alpha = 0$ for impact parameters that
lie outside $x = \lambda$ (where the density takes exactly the background
value). Thus we note, (1) for compensated lenses, lensing effects occur only
for rays that traverse the lens itself and its compensating region; (2)
despite the negative values for $\sigma$ at some radii in such a compensated
lens, the deflection angle is always positive.

Collecting formulae resulting from (\ref{eqn:sc1},\ref{eqn:m}) and (\ref
{eqn:m1}-\ref{eqn:m4}), we have that for a spherically symmetric top-hat
matter distribution, 
\begin{equation}
\alpha(x) = A\,{\frac{f(x) }{x}},  \label{eq:bend}
\end{equation}
where the constant is 
\begin{equation}
A = \left({\frac{16\pi G \rho_0 }{3c^2}} \right) (\xi_o \delta_+) \left({
\frac{\lambda^3 }{\lambda^3 -1}}\right) {\cal R}\,,  \label{eqn:const}
\end{equation}
From (\ref{eqn:39}) or (\ref{eqn:F5}), cross-overs occur where 
\begin{equation}
B(x) : = {\frac{\alpha (x) }{x}} = 1
\end{equation}
and from (\ref{eqn:42}) or (\ref{eqn:F8}) caustics occur where 
\begin{equation}
d\alpha(x)/dx = 1 \,,
\end{equation}
The maximum bending angle $\alpha_m$ occurs where $d\alpha/dx = 0$. 

Consequently,

(1) the bending angle $\alpha(x)$ is a continuous positive odd function with 
$\alpha(0) = 0$, $d\alpha/dx(0) = (3A/2) \left({\frac{\lambda^2-1 }{\lambda^2
}}\right)$, a single maximum value $\alpha_m$ at $x_m < \xi_0$ where $x_m^4
= {\frac{3 \lambda^4 (\lambda^2-1) }{4 (\lambda^6-1)}}$, and junction values 
$\alpha(1) = A (\lambda^2-1)^{3/2}/\lambda^3$, $\alpha(\lambda) = 0 =
d\alpha/dx(\lambda)$.

(2) its slope $d\alpha (x)/dx$ is an even continuous function with maximum
value $d\alpha (0)/dx=(3A/2\lambda ^2)(\lambda ^2-1)$ at the centre,
positive from $x=0$ to $x_m$, negative from $x=x_m$ to $\lambda $, and zero
thereafter, with junction values $d\alpha (1)/dx=-(A/\lambda ^3)(\lambda
^2+2)\sqrt{\lambda ^2-1}$ (here it takes its minimum value and its
derivative $d/dx(d\alpha /dx)$ is discontinuous, diverging from the left but
finite on the right) and $d\alpha (\lambda )/dx=0$. Hence caustics occur iff 
\begin{equation}
(3A/2\lambda ^2)(\lambda ^2-1)\equiv A_{crit}>1  \label{eqn:crit}
\end{equation}
(with a degenerate case when equality occurs). If they occur, say at $x=x_2$
, then $0<x_2<1$ and $x_2$ satisfies 
\begin{equation}
{\frac{d\alpha (x_2)}{dx}}={\frac A{x_2^2}}\left[ (1+2x_2^2)\sqrt{1-x_2^2}
-(1+{\frac{2x_2^2}{\lambda ^2}})\sqrt{1-{\frac{x_2^2}{\lambda ^2}}}\right]
=1\,,
\end{equation}
with the corresponding angle $\theta _2$ given by $\theta _2=x_2\xi_0/D_d$.

(3) The function $B(x)=\alpha (x)/x=Af(x)/x^2$ (with $f(x)$ given by (\ref
{eqn:m2}-\ref{eqn:m4})) is even, positive, monotonic decreasing, and
continuous, with a maximum value $B(0)=(3A/2\lambda ^2)(\lambda ^2-1)$ at
the centre, and junction values $B(1)=(A/\lambda )(\lambda
^2-1)^{3/2}/\lambda \,$, $B(\lambda )=0$. Hence cross-overs also occur iff (
\ref{eqn:crit}) is satisfied. They can occur for any value of $x>x_2$ up to $
\lambda $. If they occur, say at $x=x_1$, then 
\begin{equation}
Af(x_1)/x_1^2=1\,.
\end{equation}
with the corresponding angle $\theta _1$ given by $\theta _1=x_1\xi_0/D_d$,
where $f(x)$ is given either by (\ref{eqn:m2}) (if $x_1<1$) or by (\ref
{eqn:m3}) (if $1<x_1<\lambda $) (one cannot tell {\it a priori} in which
range it will lie; one has to try to solve one, and if there is no solution,
solve the other). Then $\theta _2(z_d,z_s)$ is the angle determining how
large a part of the sky is covered by the Einstein circle corresponding to
the cross-over surface $z=z_s$ for lenses at $z_d$ (giving multiple images
of the central point at $z_s$). 
How this scales with $z_d$ (for given $z_s$) depends on how $\rho _0$, $\xi
_0$, $\delta _{+}$, $\lambda $ and ${\cal R}$ scale with $z_d$.

(4) Pointwise over the image, the area shrinking factor is given by $\beta
= |det\,J|$ given by (\ref{eqn:38}). This can be evaluated from the formulae
given above. Using the expansion (\ref{eqn:exp}) one can evaluate this 
determinant near
the centre-line $\theta =0$; the result is 
\begin{equation}
det\,J=\left( 1-A_{crit}\right) +O(x^2)
\end{equation}
which is $1$ near the lens (when $A$ is small) and goes to $-A_{crit}$ 
(see (\ref{eqn:crit})) when $
A$ is large (the minus sign because images are reversed).

We can determine a value for the lensing parameter $M_0$ either by directly
estimating the quantities in the definition (\ref{eqn:const}), or by
estimating the maximum bending angle $\alpha_m$ for lenses considered. For
example, if $\lambda = 2$, the r.h.s. of (\ref{eq:bend}) has a maximum value
of $0.70 A$ (when $x = 0.87$). From (\ref{eqn:const}) with the bending angle
relation (\ref{eq:bend}) and angle scaling relation (\ref{eqn:10b}) we see
that then $M_0$ is determined by the relation 
\begin{equation}
M_0 \times 1.14 \times 0.7 = {\frac{2c }{H_0}} {\frac{\alpha_m }{\xi_0}} \,.
\end{equation} 
\subsection{Results}

We have written a series of Truebasic programmes that compute all the
relevant quantities for Tophat lenses, as functions of (i) the determining
parameter $A$, (ii) the source redshift $z_s$ for fixed lens redshift $z_d$,
(iii) the lens redshift $z_d$ for fixed source redshift $z_s$. We have
experimented with parameter values that correspond to observed gravitational
lensing systems; some of the results are given in the following tables and
in Figures 3 to 5. 
This area shrinking ratio is about 3 after cusps have occurred, for
scales of about 3 times the cusp scale, corresponding to the cusp image
point where the deflection is the same size as at the cusp.

\subsection{ Galaxy clusters}
We present a table of results for parameters corresponding to
four well-known {\em galaxy clusters}
that cause gravitational lensing (note that we are
not making detailed models of these objects; rather we are using their 
observed properties to determine reasonable parameter values in our 
SCL model). From the cluster Abell $2218$ (see refs. 
\cite{bi:Kneib, bi:Pello}) we have selected as images the arcs at redshifts $
z_s=2.6$ and $3.3\;$ respectively, as a case study,
where the brackets imply this is evaluated at the angle
cut-off angle. We then list the
corresponding shrinking factor $\left\langle \beta \right\rangle $ for these
two images, at the cut-off angle, 
followed by their cusp and cross-over angles. The last column is
the shrinking factor for the source placed at decoupling redshift $z_s=1200$
. We also consider other lensing clusters Abell $963$ (ref. \cite{bi:Lavery}
), Abell $370$ (refs. \cite{bi:Soucail, bi:Mell}), and Abell $2390$ (refs. 
\cite{bi:Fort, bi:Mell}).

\[
\begin{tabular}{|c|c|c|c|c|c|}
\hline
LENS & $\alpha _{max}$ & $\xi _0$ & $z_d$ & $
\begin{tabular}{c}
$z_s$%
\end{tabular}
$ & 
\begin{tabular}{c}
shrinking \\ 
$\left\langle \beta \right\rangle $%
\end{tabular}
\\ \hline
A2218 & $90^{\prime \prime }$ & $160kpc$ & $0.174$ & $3.3$/2.6 & $3.2$ \\ 
\hline
A963 & $76^{\prime \prime }$ & $130kpc$ & $0.206$ & 0.7 & $3.2$ \\ \hline
A370 & $70^{\prime \prime }$ & $100kpc$ & $0.374$ & $0.724/$1.305 & $3.1$ \\ 
\hline
A2390 & $75^{\prime \prime }$ & $160kpc$ & $0.231$ & $0.913$ & $3.3$ \\ 
\hline
\end{tabular}
\]
\[ 
\begin{tabular}{|c|c|c|c|}
\hline
LENS & 
\begin{tabular}{c}
Cusp angle \\ 
$\theta _2$%
\end{tabular}
& 
\begin{tabular}{c}
Cross-over angle \\ 
$\theta _1$%
\end{tabular}
& 
\begin{tabular}{c}
shrinking $\left\langle \beta \right\rangle $ \\ 
at decoupling
\end{tabular}
\\ \hline
A2218 & $47$/46 & $77$/75 & $3.2$ \\ \hline
A963 & 28 & 48 & $3.1$ \\ \hline
A370 & $16/$25 & $27$/36 & $3$ \\ \hline
A2390 & 24 & 41 & $3.35$ \\ \hline
\end{tabular}
\]

\subsection{Galaxies}
We have also used a set of galactic lenses, as evidenced by multiple
images of more distant objects, to provide parameters for
our model, giving the second table.
The first lens is often
referred to as the `clover leaf': $+H1413+117$ has four images of a QSO at
redshift $z_d=2.55\;$(See refs. \cite{bi:Magain, bi:Kayser}). The 
second is the seen in QSO images
A and B for the system $2345+007$ correspond to a redshift $z_d=2.15$,
despite image A being $1.7mag$ brighter than image B (ref. \cite{bi:Weedman,
bi:Steidel}). The third is the triple radio source $+MG2016+112$ 
(see ref. \cite{bi:Lawrence}). The fourth is the QSO pair in $1635+267$ with
nearly equal redshift $z_d=1.96$ \cite{bi:Djorg}. Finally a
nearly full Einstein ring was observed in $+MG1131+0456$, albeit somewhat
elliptic in shape (ref. \cite{bi:Hewitt} ).

\[
\begin{tabular}{|c|c|c|c|c|c|}
\hline
LENS & $\alpha _{max}$ & $\xi _0$ & $z_d$ & $z_s$ & 
\begin{tabular}{c}
shrinking \\ 
$\left\langle \beta \right\rangle $%
\end{tabular}
\\ \hline
+H1413+117 & $9^{\prime \prime }$ & $3kpc$ & $1.5$ & $2.55$ & 3 \\ \hline
2345+007 & $20^{\prime \prime }$ & $15kpc$ & $0.5$ & $2.15$ & 3 \\ \hline
+MG2016+112 & $30^{\prime \prime }$ & $30kpc$ & $1.01$ & $3.75$ & 3 \\ \hline
1635+267 & $30^{\prime \prime }$ & $30kpc$ & $0.57$ & $1.96$ & 3 \\ \hline 
\end{tabular}
\]

\[
\begin{tabular}{|c|c|c|c|}
\hline
LENS & 
\begin{tabular}{c}
Cusp angle \\ 
$\theta _2$%
\end{tabular}
& 
\begin{tabular}{c}
Cross-over angle \\ 
$\theta _1$%
\end{tabular}
& 
\begin{tabular}{c}
shrinking $\left\langle \beta \right\rangle $ \\ 
at decoupling
\end{tabular}
\\ \hline
+H1413+117 & $0.6$ & $0.9$ & 3 \\ \hline
2345+007 & $3$ & $6$ & 3 \\ \hline
+MG2016+112 & $5$ & $8$ & 3 \\ \hline
1635+267 & $6$ & $10$ & 3 \\ \hline
+MG1131+0456 & $0.6$ & $1.0$ & 3 \\ \hline
\end{tabular}
\]

We find that the caustics shrinking factor tends to an average factor $>3$
at large z (as required to get a 3-fold covering factor). However because of
the divergence of the light rays within the cusps, it can be much larger
for parameter values implying to strong lensing
(the actual angle corresponding to the cusps is then much smaller than in the
equivalent FL model).

\section{Conclusions}
\label{sec: concl} 

Of particular interest is the way the cusp size and the ``shrinking" vary
with redshift of the source and of the lensing object. 
This depends on two things: firstly the variation of angular sizes with
redshift, remembering (a) minimum angular apparent diameter occurs at 
$z = 1.25$, so
that the maximum angle $\theta_c$ for cusps to form due to lenses of given
size and strength will have minimum at
that redshift; and (b) that the ratio of distances that enters $
\sigma_{cr}$ saturates with increasing $z_s$ (for given $z_d$) 
but has a maximum for each $z_s$ at a $z_d$ of about $0.6$ 
which is thus the optimal distance for the lens in order to 
create cusps on the last scattering surface.
 
that are not typical of all galaxies or clusters; but they confirm 
in a concrete way the broad picture proposed in
Paper I: an area `shrinking' factor of 3 will occur for each lens
that causes cusps, on the scale of the cusps (precisely: at the cut-off angle
$\theta_3$ which gives the same deflection at the
source as the cusp angle, but on the opposite side). The total effect 
when averaging over large angular scales will 
depend on what fraction of the sky is covered by these angles for all 
lenses at all smaller scales, as a function of redshift; some simple
estimates of this overall effect were given in Paper I. 
To determine realistic multiplicity factors as a function of
redshift will require simulations with multiple lensing and more realistic 
lens models, for example standard elliptical lens models determined by 
a velocity dispersion parameter and ellipticity parameters as in 
\cite{bi:Kneib} which allow an increase in the degree of multiple covering 
(because individual elliptic lenses can have a covering factor of 5).
The effect will differ on angular scales, and will almost certainly
be substantial due to micro-lensing, with an additional increase due to 
galactic and cluster lensing. The implication of this paper and Paper I
is that it is incorrect to assume that areas average out to the 
background FL value on large angular scales; one can only know the
true area ratios - expressed in the shrinking factors considered in this
paper - by detailed calculation.


\section*{ Acknowledgments}

We thank the FRD for financial support, and T Gebbie and M Shedden for
checking the calculations.


\newpage

\begin{figure}
\epsfxsize = 5in
\epsffile{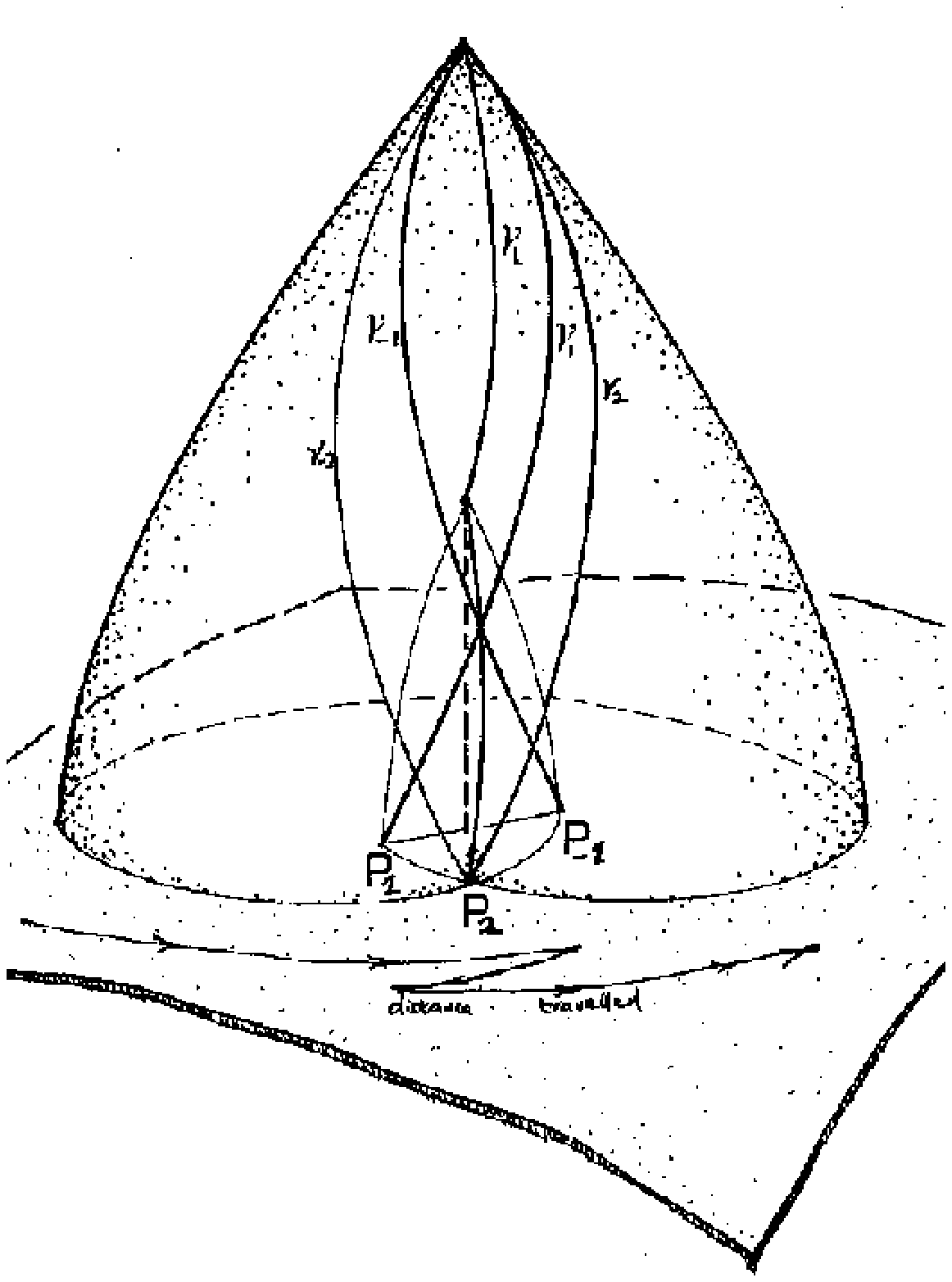}
\caption{ Shape of caustics in past light cone showing preferred
geodesics and distance traveled.
}
\label{fig:se1}  
\end{figure}

\begin{figure}
\epsfxsize = 5in
\epsffile{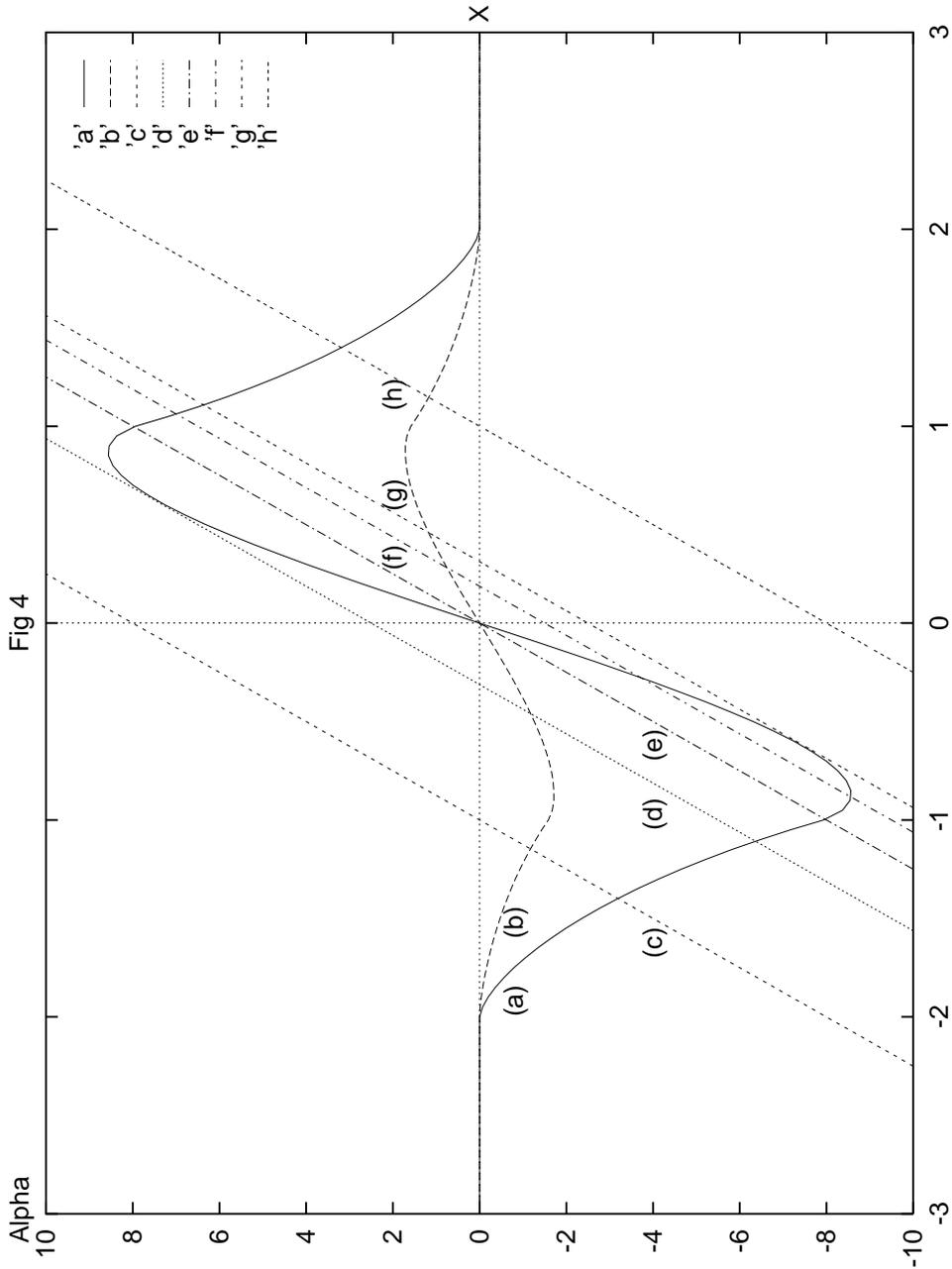}
\caption{ 
The bending angle diagram for two different redshifts. 
(a) one large, so other side of conjugate point Q; (b) One
small, so this side of conjugate point Q. 
The number of images is the number of times the line y
= x intersects the bending angle curve. Considering curve (a),
firstly, there is one image corresponding to line (c); then there are
two images for line (d) which is tangent to the curve and determines
the cusp angle; there are three 
intersections for line (e) which determines the cross-over angles 
(as it corresponds to no displacement at the source plane); there are 3 
images for generic position (f), again two images for line (g)
as it passes through the cusp, and finally one image for line
(h). Parameters based on the lens +MG1131+0456.}
\end{figure}

\begin{figure}
\epsfxsize = 5in
\epsffile{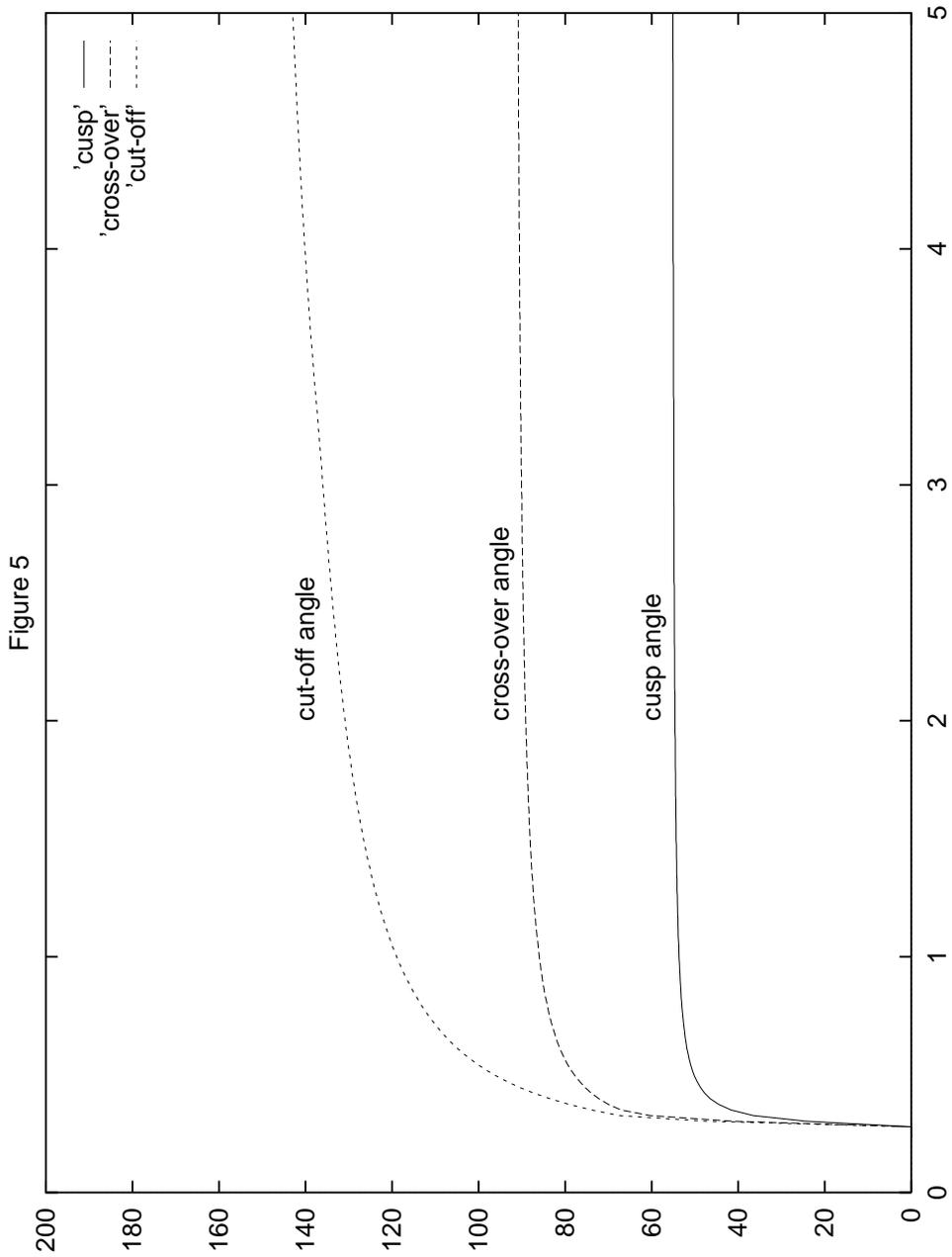}
\caption{ Variation in (a) cusp angle, (b) the crossover angle, and (c)
the cut-off angle. The
image redshift starts at the limiting value of $z_s=z_d=0.231$, 
and increases through the arc redshift of 0.914, up to
the value $z_s=5$. Parameters based on the lens A2390.}
\end{figure}

\begin{figure}
\epsfxsize = 5in
\epsffile{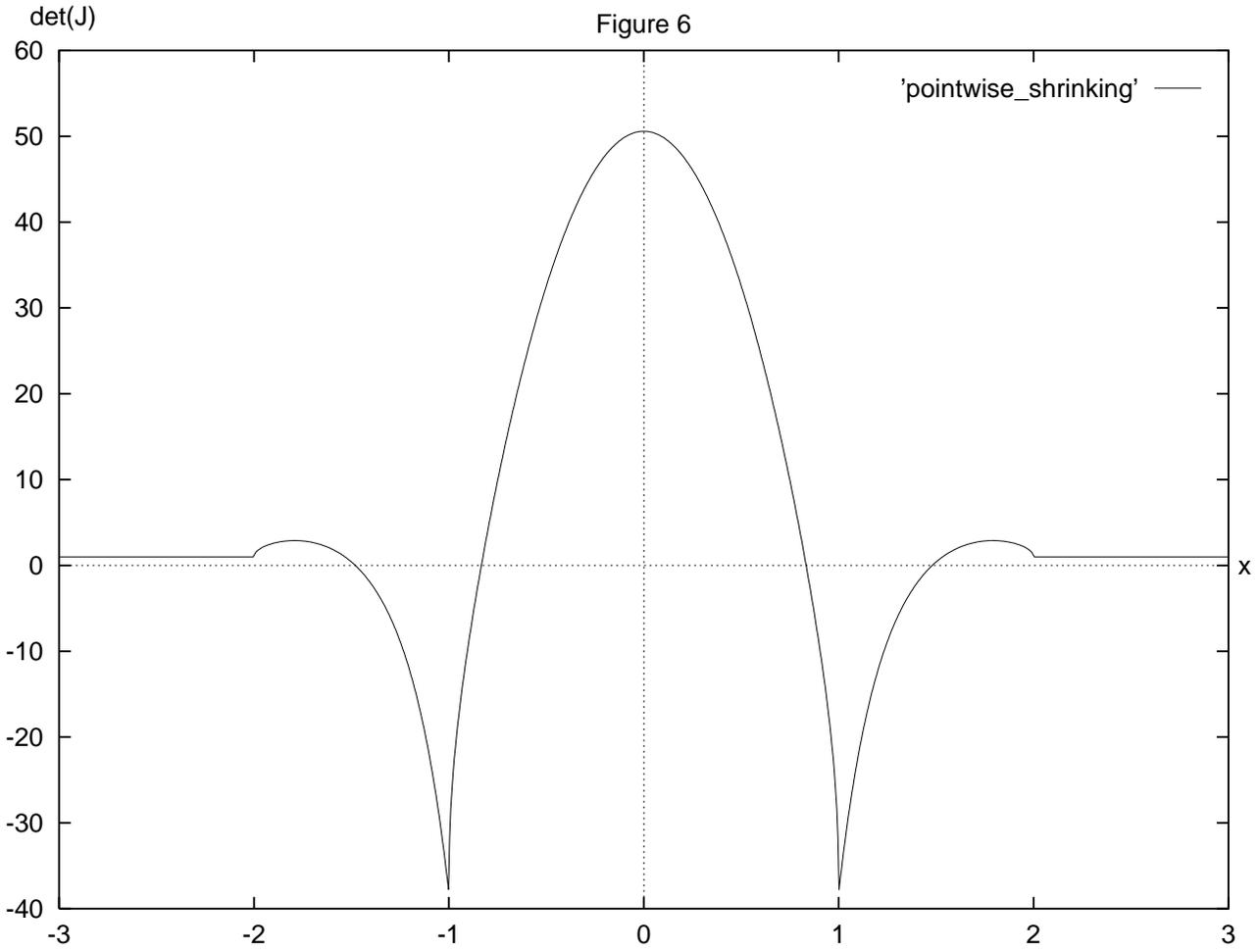}
\caption{The pointwise area shrinking ratio $\gamma $
for parameters based on the lens +MG1131+0456.}
\end{figure}

\begin{figure}
\epsfxsize = 5in
\epsffile{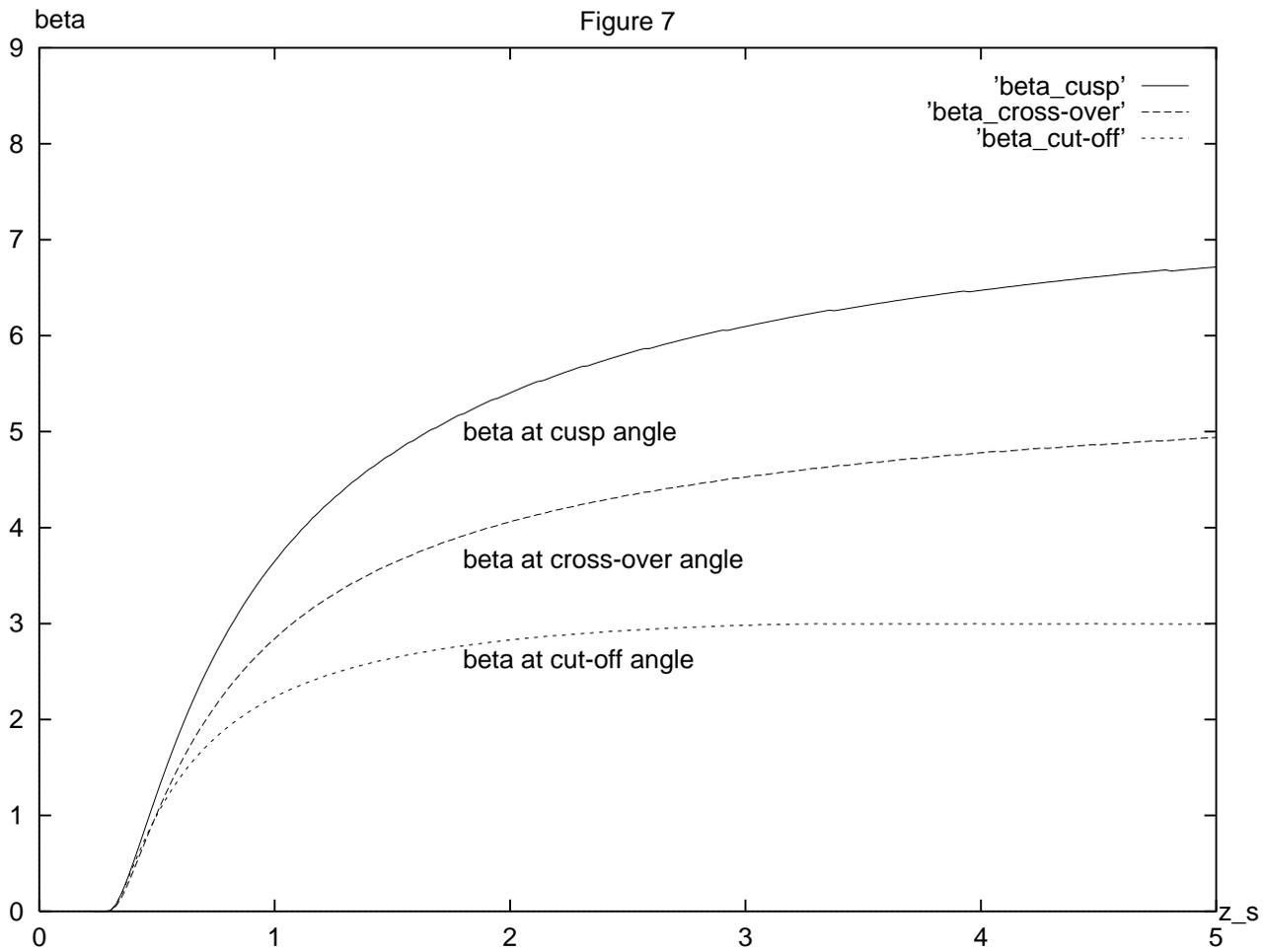}
\caption{The average
area distance shrinking ratio $\beta $ for the Abell
cluster 2390. The average $\langle \beta \rangle$  at (a) $\theta = \theta_1$
(the cusp angle), (b) $\theta = \theta_2$ (the cross-over angle), and (c)
$\theta = \theta_3$ (the cut-off angle).}
\end{figure} 

\end{document}